\documentclass[12pt,a4paper]{article}
\usepackage{amssymb}
\usepackage{epsf}
\usepackage{cite}
\newlength{\HFPP}       \HFPP5.4mm
\makeatletter
\@addtoreset{equation}{section}
\makeatother

\def\preprint#1#2{\noindent\hbox{#1}\hfill\hbox{#2}\vskip 10pt}

\begin{document}
\begin{titlepage}
\def\thefootnote{\fnsymbol{footnote}}

\preprint{ITP-UH-08/99}{March 1999, revised May 1999}
\vfill

\begin{center}
  {\Large\sc Tunneling singularities in the open Hubbard chain}
\vfill

{\sc Gerald Bed\"urftig}
	and
{\sc Holger Frahm}\footnote{e-mail: frahm@itp.uni-hannover.de}
\vspace{1.0em}

{\sl
  Institut f\"ur Theoretische Physik, Universit\"at Hannover\\
  D-30167~Hannover, Germany}
\end{center}
\vfill

\begin{quote}
We study singularities in the $I$-$V$ characteristics for sequential
tunneling from resonant localized levels (e.g.\ a quantum dot) into a
one dimensional electron system described by a Hubbard model.
Boundary conformal field theory together with the exact solution of
the Hubbard model subject to boundary fields allows to compute the
exponents describing the singularity arising when the energy of the
local level is tuned through the Fermi energy of the wire as a
function of electron density and magnetic field.  For boundary
potentials with bound states a sequence of such singularities can be
observed.
\end{quote}

{PACS-Nos.: 
05.70.Jk,	
71.10.Fd,	
71.10.Pm,	
73.20.-r	
}


\vspace*{\fill}
\setcounter{footnote}{0}
\end{titlepage}

\section{Introduction}
Electronic correlations together with strong quantum fluctuations are
known to determine the low temperature properties of quasi one
dimensional conductors.  Theoretical investigations using integrable
lattice realizations of these Tomonaga-Luttinger liquids (TLL)
together with numerical studies and field theory approaches such as
bosonization have provided much of the insight into the peculiar
properties of such systems.
Experimental evidence for TLL behaviour on the other hand is still
rare in spite of the tremendous progress in synthetization of quasi
one dimensional materials or fabrication of nano-structures in which
the transport of electrons is confined to a single one dimensional
channel \cite{LowD96,NW97,MSS8}.  One reason for this is that the
theoretical work on TLLs has concentrated on low energy bulk
properties of perfect infinitely long system which are difficult to
access experimentically.
Recently, studies of the response of a TLL on local perturbations have
become feasible due to the construction of integrable models with open
boundaries and a better understanding of quantum field theories in the
presence of a boundary.  Local inhomogeneities may have a profound
effect on the transport properties of one dimensional interacting
electron gases, even leading to phases with vanishing transmission of
a barrier \cite{kafi:92,funa:93a,funa:93b,woaf:94,sasc:95}.
Finite size effects in the resulting open chains have been studied to
understand the possible experimental consequences of TLL properties in
such systems (see e.g.\ \cite{FaGo95,Egg95,frzv:97a,Egg97,Matt97,dmrg:98}).
Among possible consequences of local perturbations are Fermi edge
singularities which may be observed in X-ray absorption amplitudes and
--- as will be discussed in this paper --- in tunneling experiments
\cite{Calleja91,Geim94,Smoliner96,LiangX97}.  In both cases the nature
of the singularities is strongly affected by the properties of the
TLL.

In this paper we study tunneling from a resonant localized level into
a TLL in the limit of low barrier conductance as observed in the
current-voltage characteristic at zero bias.  As a specific
representation of the latter we choose the one-dimensional Hubbard
model.  In this lattice model we find --- similar as in previous work
on optical absorption processes \cite{esfr:97} --- a rich spectrum of
edge singularities due to the existence of bound states reflecting the
separation of spin and charge in the TLL \cite{befr:97}.
An experimental realization of the tunneling processes under
investigation might be a quantum dot (providing the localized level)
coupled to a quantum wire.  The reservoir supplying charges to fill
the state in the quantum dot is left unspecified.  The energy $E_i$ of
the local level can be tuned by varying a gate voltage of the dot.
The only influence of the quantum dot on the Luttinger liquid
considered below is the electrostatic interaction with its net charge
(we consider the dot occupied with a single electron to be
electrically neutral).  This description is very similar to the model
invoked to describe the X-ray edge singularity in metallic systems
\cite{nodo:69} and has previously been used to study tunneling from a
resonant local level into two- and three-dimensional systems
\cite{mala:92}.  These considerations lead to the following
Hamiltonian
\begin{eqnarray}
{\cal H}=&-& \sum_{\sigma,j=1}^{L-1}
       \left( c_{j,\sigma}^\dagger c_{j+1,\sigma}+h.c.\right)
       +4u \sum_{j=1}^{L}n_{j \uparrow}n_{j \downarrow}+\mu \hat{N}
       -{h \over 2} (\hat{N}-2\hat{N}_\downarrow) \nonumber \\
      &+& E_ib^\dagger b- bb^\dagger
      	p (\hat{N}_{1,\uparrow}+\hat{N}_{1,\downarrow})\ . 
\label{ham:hubb}
\end{eqnarray}
where $b^\dagger$ ($b$) are canonical fermionic creation
(annihilation) operators for a spin-$\uparrow$ electron the localized
state and $c_{j\sigma}^\dagger$ creates an electron of spin $\sigma$
on site $j$ of the one-dimensional chain.  The chemical potential
$\mu$ and magnetic field $h=g\mu_B H$ allow to control the filling
factor and magnetization of the quantum wire.
Upon variation of the gate voltage tunneling between the local level
and the wire becomes possible if the energy $E_i$ of the local level
exceeds the Fermi energy.  We restrict ourselves to the case where the
barrier conductance is low.  Hence, the transport is dominated by
incoherent sequential tunneling processes and we can neglect Coulomb
blockade effects and higher order processes such as ``cotunneling''
\cite{SCT92}.  Within the ``orthodox theory'' \cite{AL91} the current
due to sequential tunneling is computed by application of the golden
rule leading to
\begin{equation}
I(E_i) \propto \sum_n |\langle n | c_{1,\uparrow}^\dagger b 
|\tilde{O} \rangle |^2 \delta(E_n-E_0-E_i)\ .
\label{eq:ie}
\end {equation}
Here $|\tilde O\rangle = b^\dagger |0\rangle$ denotes the ground state
of the open Hubbard chain in the $N_e$-particle sector with the local
level occupied and hence vanishing boundary potential $p$.  
The sum in (\ref{eq:ie}) extends over all eigenstates $|n\rangle$ of
the chain in the $(N_e+1)$-particle sector in the presence of the
boundary potential $p$.  Eq.~(\ref{eq:ie}) can be rewritten as a
Fourier integral:
\begin{equation}
I(E_i) \propto \mbox{Re} \int_0^\infty {\rm d}t\ e^{i E_i^+ t} 
\langle \tilde{0} | b^\dagger(t)c_{1,\uparrow}(t)c_{1,\uparrow}^\dagger(0) b(0)|\tilde{0}\rangle \,
\label{eq:iei}
\end{equation}
where $E_i^+=E_i+i0$.  Near the threshold $E_i \approx E_{th}$ the
intensity exhibits a characteristic singularity:
\begin{equation}
  I(E_i)\propto {1 \over |E_i-E_{th}|^\alpha}\ .
\label{eq:ea}
\end{equation}
For non interacting electrons the exponent $\alpha$ can be expressed
in terms of the phase shift at the Fermi surface \cite{mala:92}.  As
in the case of the X-ray edge singularity one expects several
thresholds if the electrostatic potential $p$ is strong enough to form
bound states in the TLL \cite{cono,affl:97}.  In this paper we want to
study this problem for tunneling into a TLL where an additional
dependence of the exponent $\alpha$ on the interaction parameters
(i.e.\ electron density, magnetization and strength of the Hubbard
interaction $4u$) in (\ref{ham:hubb}) of the Luttinger liquid is to be
expected from the results obtained for the related X-ray problem (see
Refs.~\cite{ogfn:92,aflu:94,esfr:97}).  In the following section we summarize
the relevant properties of the model (\ref{ham:hubb}) obtained from
its Bethe Ansatz solution.  From this solution combined with results
from boundary conformal field theory (BCFT)
\cite{card:89,aflu:94,affl:94} we extract the spectrum of thresholds
and the corresponding exponents $\alpha$.
\section{Bethe Ansatz Solution of the model}
The Bethe Ansatz equations (BAE) determining the spectrum of ${\cal
H}$ with empty local state (i.e.\ with boundary chemical potential
$p$) in the $N_{e}$-particle sector with magnetization $M={1\over2}
N_{e}-N_{\downarrow}$ read \cite{assu:96,deyu:pp,shwa:97}:
\begin{eqnarray}
  {\rm e}^{ik_j (2L+1)} B_c(k_j) = 
  \prod_{\beta=-N_\downarrow}^{N_\downarrow}
  \frac{\sin k_j-\lambda_\beta+iu}{\sin k_j-\lambda_\beta-iu}, \quad  
  j=-N_e,\ldots,N_e \nonumber
  \\ \nonumber \\
  B_s(\lambda_\alpha) \prod_{j=-N_e}^{N_e}  
  \frac{\lambda_\alpha-\sin k_j+iu}{\lambda_\alpha-\sin k_j-iu} = 
  \prod_{\beta=-N_\downarrow \atop \beta \not= \alpha}^{N_\downarrow}  
  \frac{\lambda_\alpha-\lambda_\beta+2iu}{\lambda_\alpha-\lambda_\beta-2iu},
  \quad \alpha=-{N_\downarrow},\ldots,{N_\downarrow}  
   \label{eq:bae}
\end{eqnarray}
where one should identify $k_{-j}\equiv-k_j$ and $\lambda_{-\alpha}
\equiv-\lambda_{\alpha}$.  The boundary phase shifts appearing in the
BAE read
\begin{equation}
  B_c(k)=\left(\frac{{\rm e}^{ik}-p}{1-p {\rm e}^{ik}}\right)
  \frac{\sin k +iu}{\sin k -iu}, \qquad  B_s(\lambda)= 
  \frac{\lambda+2iu}{\lambda-2iu}\ .
  \label{bound}
\end{equation}
The energy of the eigenstate of Eq.\ (\ref{ham:hubb}) corresponding to
a solution of the BAE is given by
\begin{equation}
  E= \sum_{j=1}^{N_e}\left({\mu} - {h \over 2}-2\cos k_j\right)
   + h N_\downarrow\ .
  \label{energy}
\end{equation}
In Refs.~\cite{assu:96,deyu:pp,shwa:97} the ground state and the
low--lying excitations of this model where studied for small boundary
fields.  In \cite{befr:97} the existence of boundary states for $|p|>1$
has been established.  In the Bethe Ansatz solution these bound states
manifest themselves as additional complex solutions for the charge and
spin rapidities.  In Fig.~\ref{fig:enpm} the spectrum of bound states
for $u=1$ is shown.
Using standard procedures, the BAE for the ground state and low--lying
excitations in the thermodynamic limit can be rewritten as linear
integral equations for the densities $\rho_c(k)$ and $\rho_s(\lambda)$
of real quasi-momenta $k_j$ and spin rapidities $\lambda_\alpha$,
respectively:
\begin{equation}
   \left( \begin{array}{c} \rho_c(k) \\[5pt] 
                           \rho_s(\lambda) \end{array} \right) =
    \left( \begin{array}{c} {1 \over \pi}+{1 \over L}\hat{\rho}_c^0(k)\\[5pt] 
                       {1 \over L}\hat{\rho}_s^0(\lambda) \end{array} \right)
    + K * \left( \begin{array}{c} \rho_c(k') \\[5pt] 
                           \rho_s(\lambda') \end{array} \right)\ 
\label{eq:dnorm}
\end{equation}
with the kernel $K$ given by
\begin{equation}
 K=
   \left( \begin{array}{cc} 0 & \cos k\ a_{2u}(\sin k -\lambda') \\[5pt] 
                            a_{2u}(\lambda-\sin k') & 
                -a_{4u}(\lambda-\lambda') \end{array} \right).
 \label{eq:kernel}
\end{equation}
Here we have introduced $a_y(x)={1 \over 2\pi}\frac{y}{y^2/4+x^2}$,
and $f*g$ denotes the convolution $\int_{-A}^{A}{\rm d}y f(x-y)g(y)$ with
boundaries $A=k^{(0)}$ in the charge and $A=\lambda^{(0)}$ in the spin
sector.  These boundaries are functions of the external chemical
potential $\mu$ and magnetic field $h$.  Alternatively, in a canonical
approach the values of $k^{(0)}$ and $\lambda^{(0)}$ are fixed by the
conditions
\begin{equation}
   \int_{-k^{(0)}}^{k^{(0)}}{\rm d}k \rho_c =
        \frac{2\left[N_e-C_c\right]+1}{L}, \qquad
   \int_{-\lambda^{(0)}}^{\lambda^{(0)}}{\rm d}\lambda\rho_s =
        \frac{2\left[N_\downarrow-C_s\right]+1}{L} ,
\label{eq:fix}
\end{equation}
where $C_c$ ($C_s$) denotes the number of complex $k$
($\lambda$)--solutions present in the ground state \cite{befr:97}.
The boundary phase shifts (\ref{bound}) and the presence of complex
solutions to the BAE determines the driving terms $\hat{\rho}_c^0$ and
$\hat{\rho}_s^0$ in (\ref{eq:dnorm}).  Their explicit form can be
found in Refs.~\cite{assu:96,deyu:pp,shwa:97,befr:97}.  Denoting the
solutions of (\ref{eq:dnorm}) \emph{without} the constant contribution
$1/\pi$ to the driving term by $\hat{\rho}_c$ and $\hat{\rho}_s$ we
introduce shift angles
\begin{equation}
  \theta^c_p = {1 \over 2}\left(L 
	\int_{-k^{(0)}}^{k^{(0)}}{\rm d}k \hat{\rho}_c-1+2
    C_c \right), \qquad
\theta^s_p = {1 \over 2}\left(L 
	\int_{-\lambda^{(0)}}^{\lambda^{(0)}}{\rm d}\lambda 
  \hat{\rho}_s -1+2 C_s \right) .
\label{eq:tpc}
\end{equation}
Following Woynarovich \cite{woyn:89} one can calculate the finite size
spectrum of the model, reproducing the result of \cite{assu:96}:
\begin{eqnarray}
  E = L e_\infty + f_\infty &+&
        {\pi v_c \over{L}} \left\{ -{1\over24} +
        {1\over{2\det^2({\bf\sf Z})}} 
                \left[\left(\Delta N_c^0 -\theta^c_p\right)Z_{ss}
		- \left(\Delta N_s^0 -\theta^s_p\right)Z_{cs}\right]^2
		+N_c^+\right\}
\nonumber \\
        &+& {\pi v_s \over{L}}\left\{-{1\over24}
        +{1\over{2\det^2({\bf\sf Z})}} 
		\left[\left(\Delta N_s^0 -\theta^s_p\right)Z_{cc}
		- \left(\Delta N_c^0 -\theta^c_p\right)Z_{sc}\right]^2
		+N_s^+\right\}. \nonumber \\
\label{fsopen:h}
\end{eqnarray}
Here $L e_\infty$ and $f_\infty$ denote the bulk and boundary energy,
$N_{c,s}^+$ are non negative integers counting the number of particle
hole excitations at the Fermi points and the $v_{c,s}$ are the Fermi
velocities of the massless charge and magnetic modes.  $\Delta
N_{c,s}$ specify the quasi-particle content of the state, in a TLL
these are holons/anti-holons in the charge sector and spinons in the
magnetic sector of the theory.
The dressed charge matrix ${\bf\sf Z}$ \cite{woyn:89,frko:90,frko:91}
\begin{equation}
{\bf\sf Z}= \left( \begin{array}{cc} Z_{cc} & Z_{cs} \\ 
  Z_{sc} & Z_{ss} \end{array} \right)=
  \left( \begin{array}{cc} \xi_{cc}(k^{(0)}) & \xi_{sc} (k^{(0)}) \\ 
  \xi_{cs} (\lambda^{(0)}) & \xi_{ss}(\lambda^{(0)}) \end{array} \right)^\top
\label{eq:z}
\end{equation}
is defined in terms of the integral equation
\begin{equation}
\left( \begin{array}{cc} \xi_{cc}(k) & \xi_{sc}(k) \\ 
  \xi_{cs}(\lambda) & \xi_{ss}(\lambda) \end{array} \right)=
\left( \begin{array}{cc} 1 & 0 \\ 0 & 1 \end{array} \right)+
        K^\top *
 \left( \begin{array}{cc} \xi_{cc}(k') & \xi_{sc}(k') \\ 
  \xi_{cs}(\lambda') & \xi_{ss}(\lambda') \end{array} \right) \,. \ 
\end{equation}
\section{Tunnel exponents}
Results from boundary conformal field theory allow to extract the
exponent $\alpha$ in Eq.~(\ref{eq:ea}) from the finite size spectrum
(\ref{fsopen:h})\cite{card:89,aflu:94,affl:94}: the Green's function
of an operator ${\cal O}$ with dimension $x$ on the complex half plane
is given by:
\begin{equation}
  \langle A | {\cal O}(\tau_1){\cal O}^\dagger(\tau_2)|A\rangle=
  {1 \over (\tau_1-\tau_2)^{2x}} 
\label{eq:ip}
\end{equation}
Conformal mapping of the half plane onto a strip of finite width $L$
allows to extract the scaling dimension of the boundary changing
operator ${\cal O}$ from the finite size spectrum (\ref{fsopen:h}) by
taking differences of the energy $E_A^0$ of the system's ground state
$|A\rangle$ in the $N_e$-particle sector without boundary potential
and the energy $E_B^{n}$ of the lowest excitation $|B,n\rangle$ in the
$(N_e+1)$-particle sector with boundary potential $p$ and non
vanishing form factor $|\langle B,n | {\cal O}^\dagger|A\rangle|^2$
(see Refs.\ \cite{frko:90,frko:91,esfr:97,befr:97} for details on the CFT
approach to the asymptotics of correlation functions in the Hubbard
model).
In the present problem $E_A^0$ is obtained from (\ref{fsopen:h}) by
choosing $\Delta N_\alpha^0=\theta_0^\alpha$ while for $E_B^0$ we have
to choose $\Delta N_c^0=1+\theta_0^c$ and $\Delta N_s^0=\theta_0^s$
corresponding to an extra spin-$\uparrow$ electron created by ${\cal
O}^\dagger$.  Writing the $O(L^{-1})$ terms in the energy difference
as $(\pi/L)\left(v_c x_c+v_s x_s\right)$ the corresponding edge
exponent in Eq. (\ref{eq:ea}) is given as
\begin{equation}
  \alpha = 1-2(x_c+x_s)\ .
\end{equation}

We now want to study the exponents at the several possible
thresholds. To gain some more insight into the role of the boundary
states we begin with a discussion of noninteracting fermions.
\subsection{Ferromagnetic case}
For sufficiently large magnetic field the electrons are polarized
ferromagnetically, hence an explicit expression is available for the
wave function in terms of a Slater determinant of single-particle
states.  Considering $|p|<1$ these are plane waves corresponding to
real wave numbers $k$ exist and we expect a single threshold.  The
corresponding edge exponent is a function of $p$ and the density of
electrons $n_e=N_e/L$ (see also Ref.~\cite{zago:97}):
\begin{equation}
  \alpha^r=1-2x_p^r=1-\left(1+
      {1\over\pi}\arctan\left({p+1\over p-1}\ \tan{\pi n_e\over2} \right)
      +{n_e \over 2}\right)^2\ .
\label{eq:xfk1}
\end{equation}
For $|p|>1$ the boundary potential can bind an electron which leads to
the existence of two thresholds, depending on whether the bound state
is occupied in the final state.  For an empty bound state we obtain
the result $\alpha^r$ of Eq.~(\ref{eq:xfk1}) for the edge exponent.  A
different exponent is found if the bound state occupied:
\begin{equation}
  \alpha^c=1-2x_p^c=1-\left(
      {1\over\pi}\arctan\left({p+1\over p-1}\ \tan{\pi n_e\over2} \right)
      +{n_e \over 2}\right)^2\ .
\label{eq:xfk2}
\end{equation}
These predictions can be checked by studying the finite-size behaviour
of the form factors.  From the conformal mapping mentioned above one
expects \cite{affl:97,befr:97}
\begin{equation}
  |\langle p |c_{1,\uparrow}^\dagger |0 \rangle|
\sim \left({1 \over L}\right)^{x_p^c} \quad \mbox{and} \quad
  |\langle \overline{p}|c_{1,\uparrow}^\dagger |0 \rangle|
\sim \left({1 \over L}\right)^{x_p^r}\ .
\label{eq:nsk}
\end{equation}
where $|p \rangle$ and $|\overline{p} \rangle$ denote the ground state
and the lowest state with empty bound state in the $(N_e+1)$-particle
sector with boundary potential $p>1$.
Note that exponent $x_p^r$ vanishes in the limit $p \to 0$. This
coincides with exact result, $I(E_i)\propto \sqrt{1-{E_i^2 \over 4}}
\Theta\left(E_i+2\cos(\pi n_e)\right)$, which does not exhibit a
singularity.

\begin{figure}[ht]
\begin{center}
\leavevmode
\epsfxsize=0.9\textwidth
\epsfbox{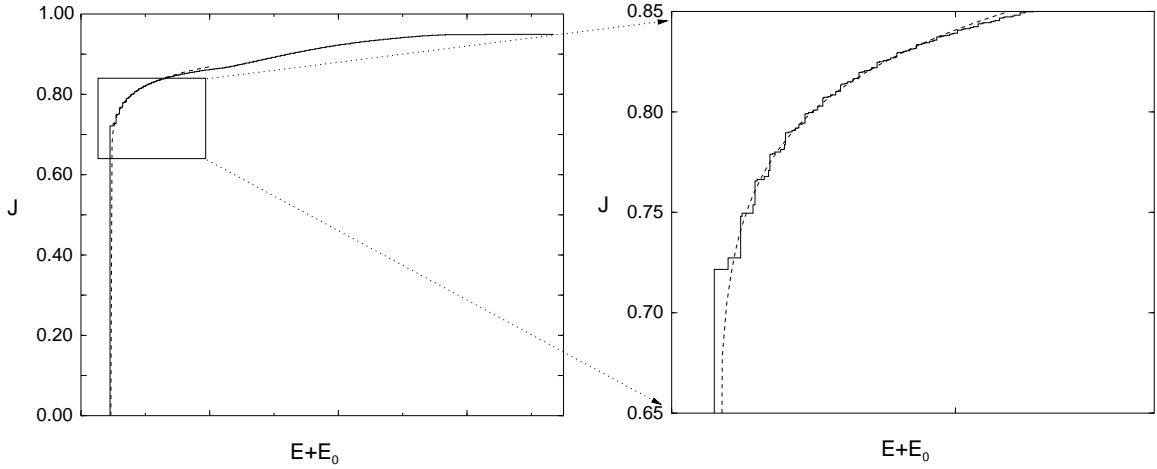}
\end{center}
\vspace{-1cm}
\caption{
\label{fig:sumk}
Numerical results for $J(E)$ for a system of $L=80$ sites with a
density $n_e=0.2$ of spin-$\uparrow$ electrons and $p=3$. The dashed
line is the fit to Eq.~(\protect{\ref{eq:fit}}).  The jump of $J(E)$
at the threshold is characteristic to a singularity with a positive
exponent $\alpha$.  Its height is given by the matrix element $\langle
p| c_{1,\uparrow}^\dagger |0 \rangle$ which vanishes in the
thermodynamic limit according to Eq.~(\protect{\ref{eq:nsk}}).  This
is the well known orthogonality catastrophe
\protect{\cite{and:67,qfyu:96}}.}
\end{figure}

\begin{figure}[ht]
\begin{center}
\leavevmode
\epsfxsize=0.9\textwidth
\epsfbox{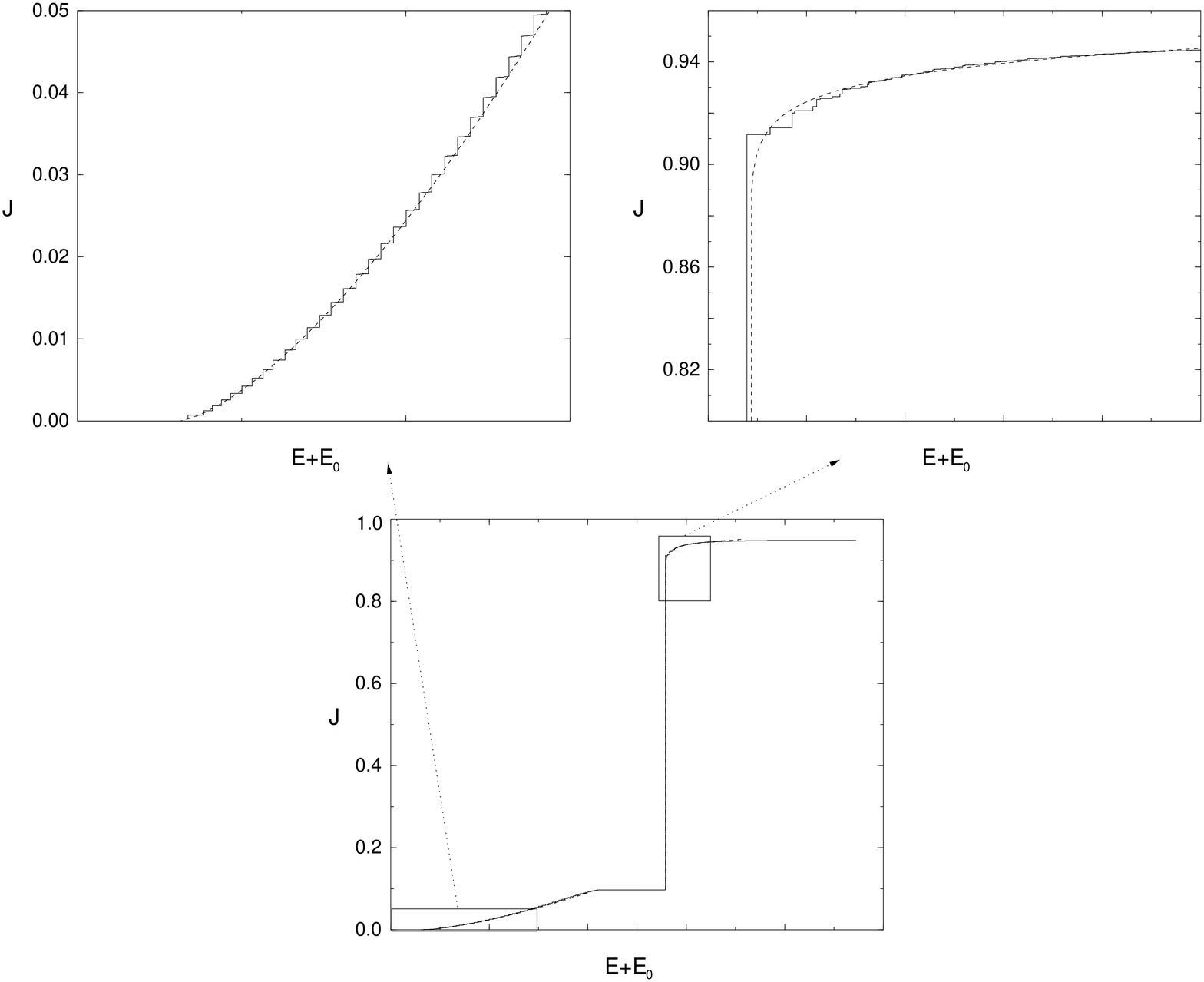}
\end{center}
\vspace{-1cm}
\caption{
\label{fig:sumkn}
Same as Fig.~\protect{\ref{fig:sumk}} but for $p=-3$.}
\end{figure}

\begin{figure}[ht]
(a)
\epsfxsize=0.45\textwidth
\epsfbox{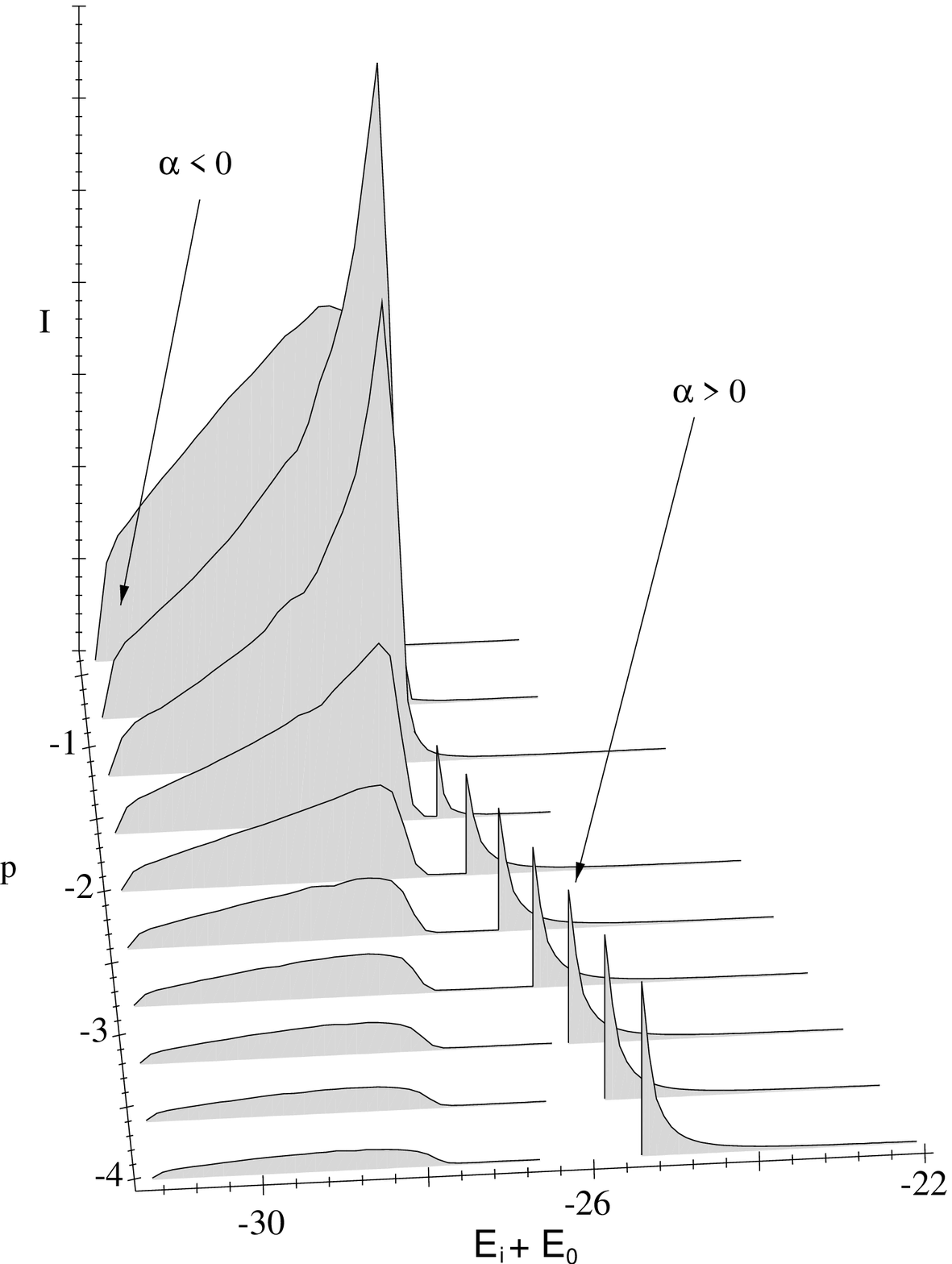}
\hfill
(b)
\epsfxsize=0.45\textwidth
\epsfbox{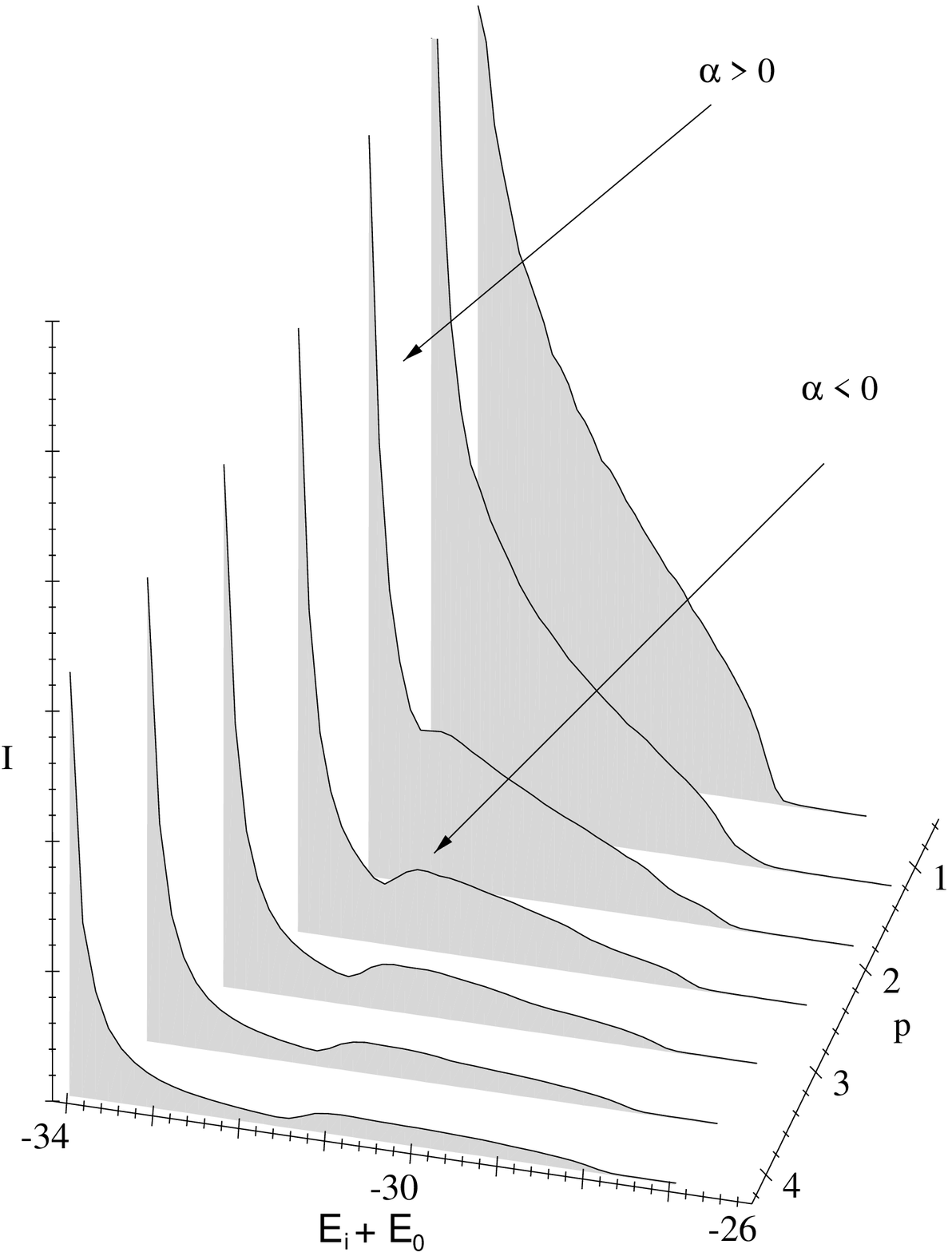}\\
\caption{
\label{fig:fnp}
Numerical results for $I(E_i)$ for systems of size $L=80$ with $16$
electrons for different $p<0$ (a) and $p>0$ (b).  The singularities
are suppressed due to the numerical differentiation.}
\end{figure}

Eq.~(\ref{eq:ie}) is now evaluated numerically.  To avoid the use of
explicit representations of the $\delta$-function we use the integral
\begin{equation}
  J(E) \equiv \int_{-\infty}^E {\rm d}E_i \ I(E_i)\ .
\label{eq:iie}
\end{equation}
Typical numerical results for $J(E)$ are shown in Figs.~\ref{fig:sumk}
and \ref{fig:sumkn}.  To make the numerical analysis of (\ref{eq:ie})
feasible we had to restrict the sum $\sum_n$ to the most important
states.  The error of this approximation has been estimated using the
sum rule $J(E \to E_{max})=1-\langle N_1 \rangle$, which is satisfied
to $>$99\% in all cases.  After smoothing of $J(E)$ and numerical
differentiation it is possible to obtain $I(E_i)$.  The result is
shown in Fig.~\ref{fig:fnp} for several potentials.
For positive $p$ the exponent is always positive at the absolute
threshold and negative at the second one for $p>1$. The situation
changes completely for $p<0$ where the ground state is always
parametrized by real wave numbers $k$ giving a negative exponent
$\alpha$ at the absolute threshold.  Occupation of the anti-bound
state corresponding to a complex $k$ leads to a positive exponent.

\begin{figure}[ht]
\begin{center}
\leavevmode
\epsfxsize=0.5\textwidth
\epsfbox{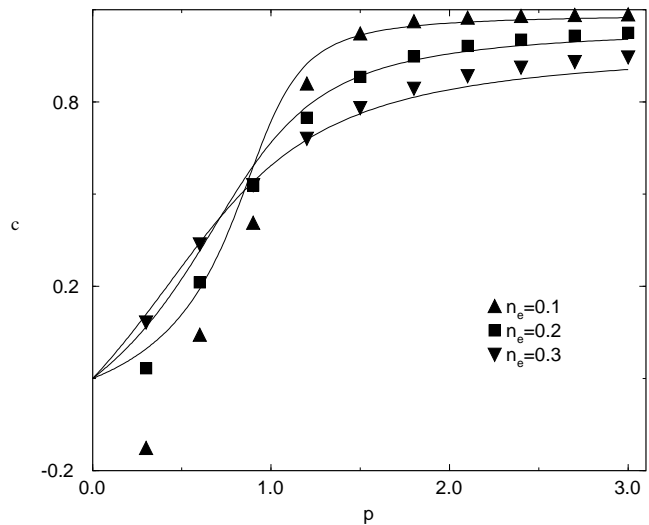}
\end{center}
\caption{
\label{fig:fitres}
Comparison of the fit (\protect{\ref{eq:fit}}) to the numerical data
with the BCFT results (solid lines) for three different electron
densities.}
\end{figure}

For the X-ray edge problem one can show that the functional dependence
of $J(E)$ is nearly unchanged by increasing the system size
\cite{dow:80,ohta:90}.  This allows to extract quantitative
informations of rather small systems ($L=80$ in the present case) by
fitting of $J(E)$ to the trial function:
\begin{equation}
   f(E)=a (E+b)^c\ .
\label{eq:fit}
\end{equation}
The resulting exponent $c$ can be compared to the BCFT-results. Using
the 600 lowest states contributing to the sum (\ref{eq:ie}) we obtain
a good agreement with the CFT-results (see Fig.~\ref{fig:fitres}) for
$p>1$ and electron densities $n_e\lessapprox0.4$.  For larger
densities a bigger discrepancy between the numerical results and the
BCFT predictions is found due to stronger finite size effects.

The second threshold due to the presence of a bound state is most
pronounced for $p<-1$.  Here the jump of $J(E)$ characteristic of a
positive edge exponent occurs at the second threshold (see
Fig.~\ref{fig:sumkn}).  While the BCFT result for the edge exponent at
the absolute threshold is $\alpha_{abs}=\alpha^r=-0.33$ the fit to the
numerical data gives $\alpha_{abs}=-0.46$ --- this indicates that a
genuine singularity is strongly affected by finite size effects.
On the other hand the numerical value for the positive exponent at the
threshold corresponding to occupied anti-bound state,
$\alpha^c=0.988$, is in very good agreement with the CFT-result
$\alpha^c=0.977$.
\subsection{Magnetic field dependence of edge exponents}
\begin{figure}[ht]
\begin{center}
\leavevmode
(a)\hspace{-6mm}
\epsfxsize=0.55\textwidth
\epsfbox{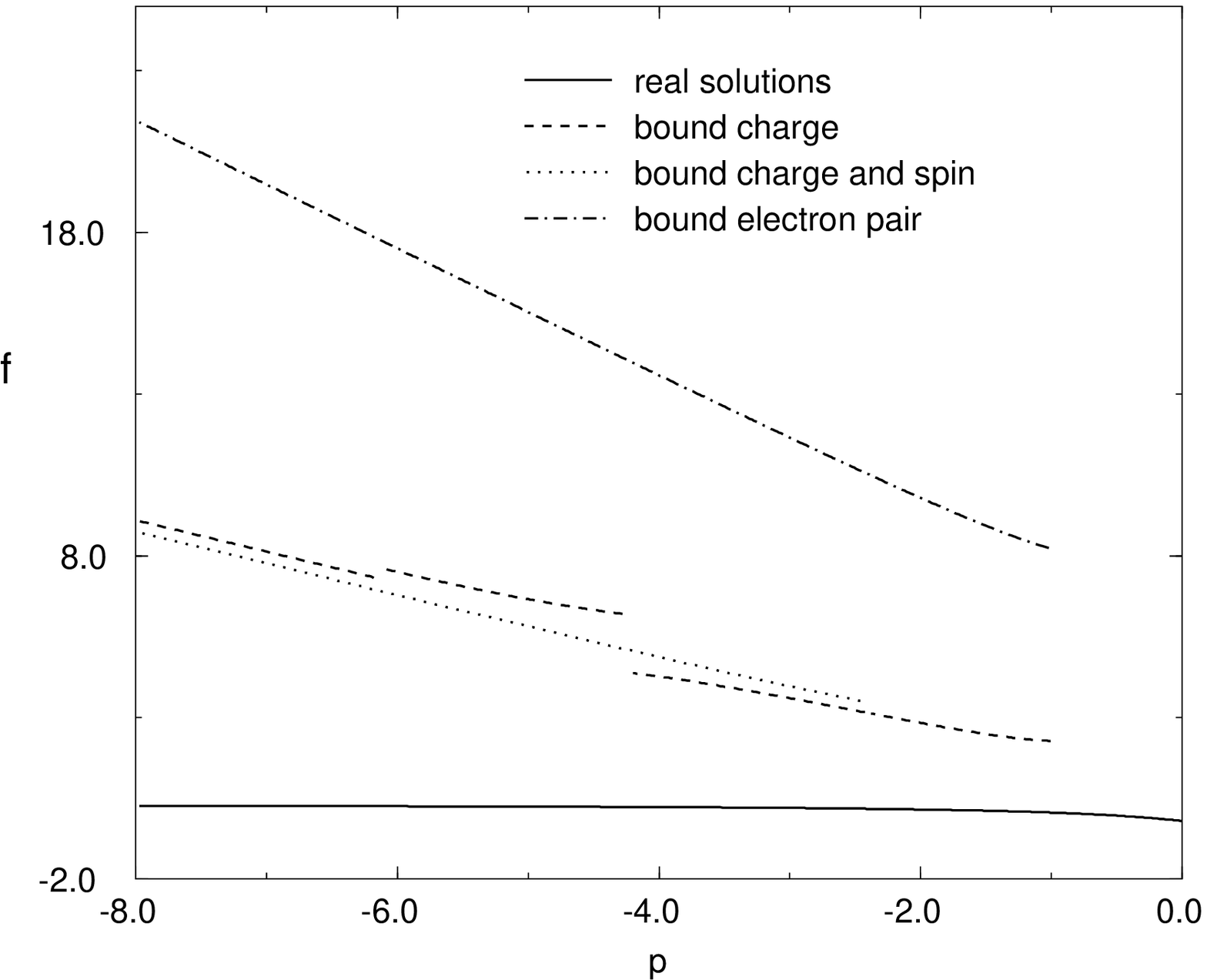}

(b)\hspace{-6mm}
\epsfxsize=0.55\textwidth
\epsfbox{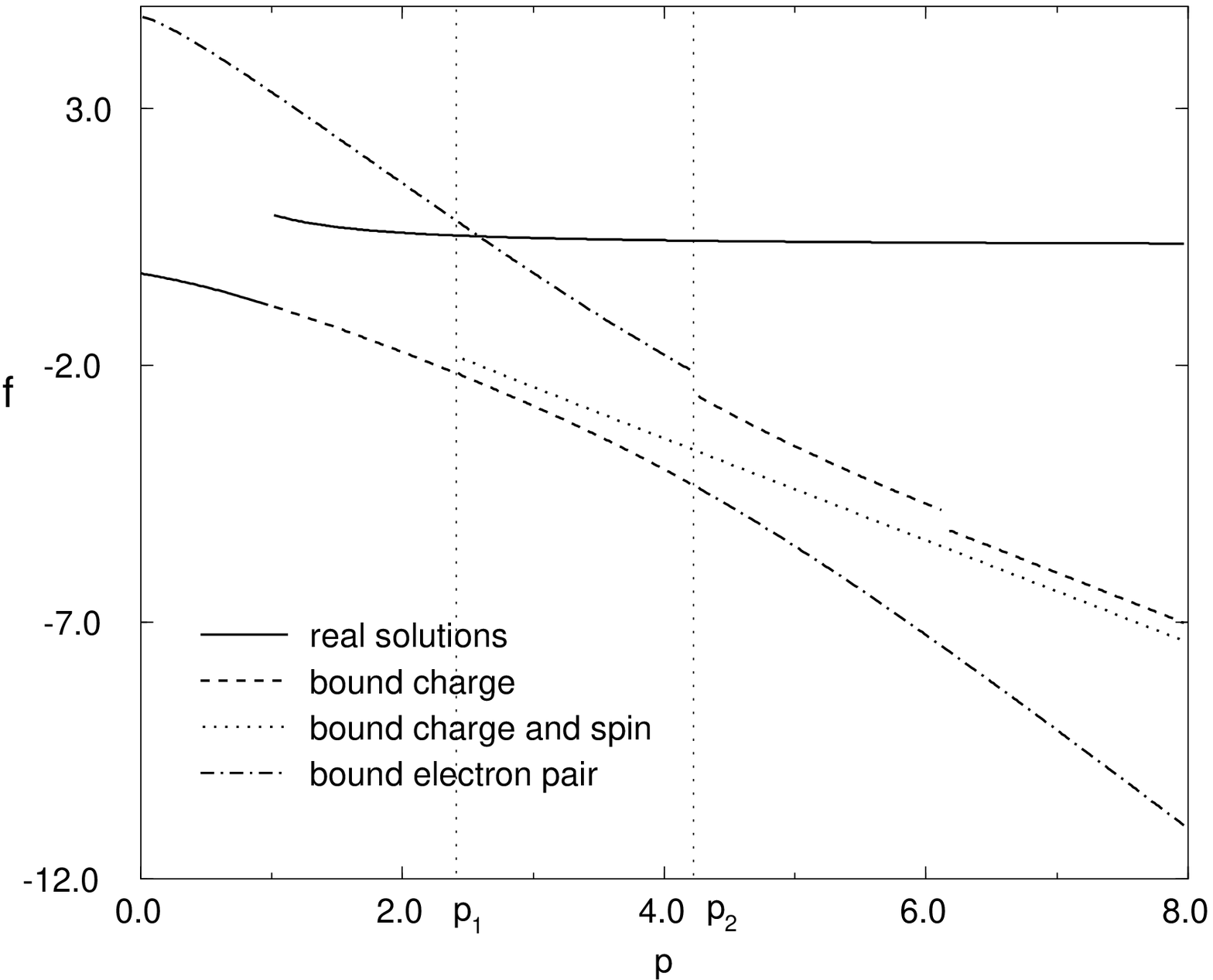}
\end{center}
\caption{ 
\label{fig:enpm}
Spectrum of boundary bound states of (\protect{\ref{ham:hubb}}) for
$u=1$ with chemical potential $\mu=0.5$ and magnetic field $h=0.3$,
for (a) $p<0$ and (b) $p>0$.  The thresholds $p_1$ for binding a
charge and spin and $p_2$ for binding of a singlet pair of electrons
are indicated.}
\end{figure}

\begin{figure}[ht]
\begin{center}
\leavevmode
\epsfxsize=0.9\textwidth
\epsfbox{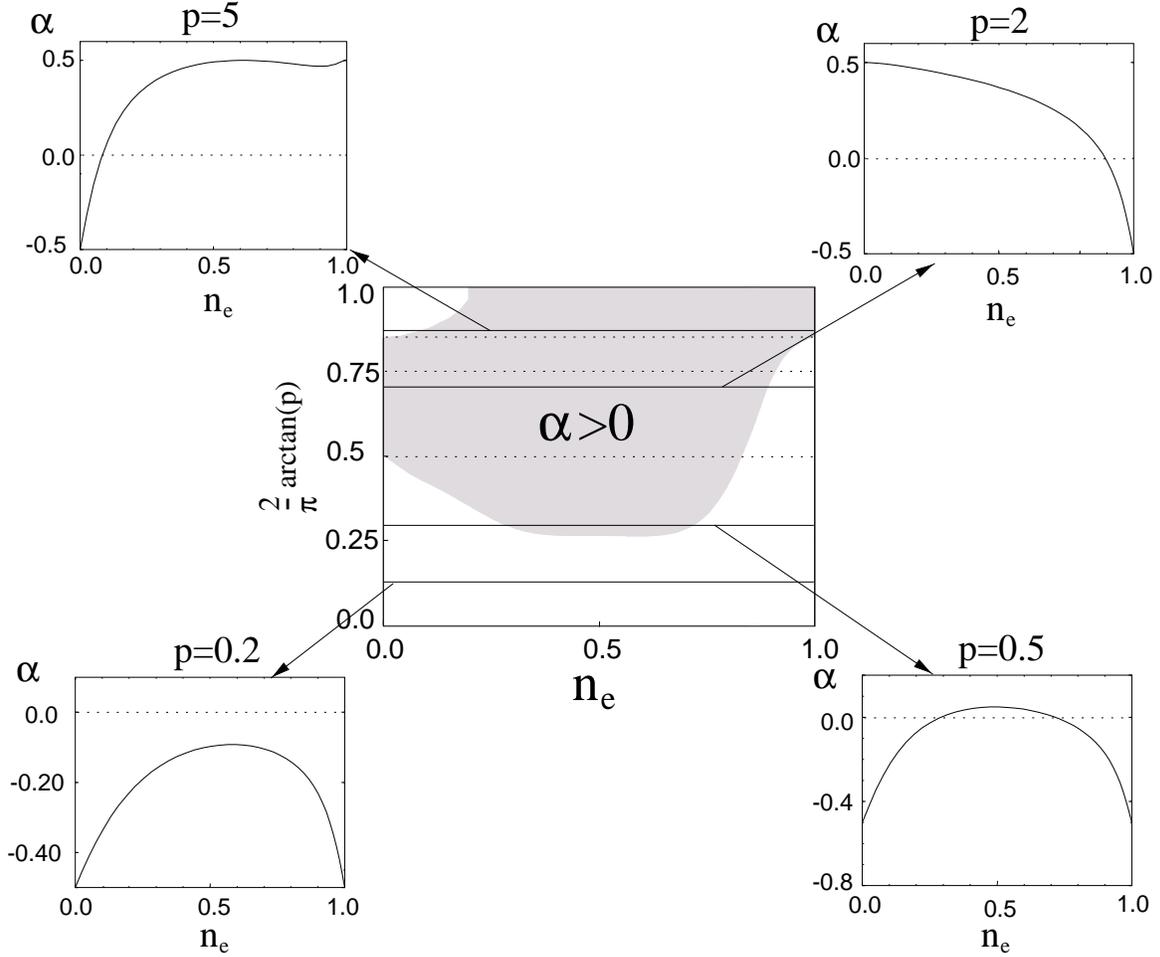}
\end{center}
\caption{\label{fig:h0a} Region with positive absolute exponent
$\alpha_{abs}$ resulting in a singularity in the $I$-$V$
characteristics for vanishing magnetic field for fixed $u=1$.  The
insets show $\alpha_{abs}$ for fixed values of $p$ as a function of
the electron density $n_e$. }
\end{figure}

\begin{figure}[ht]
\begin{center}
\leavevmode
\epsfxsize=0.7\textwidth
\epsfbox{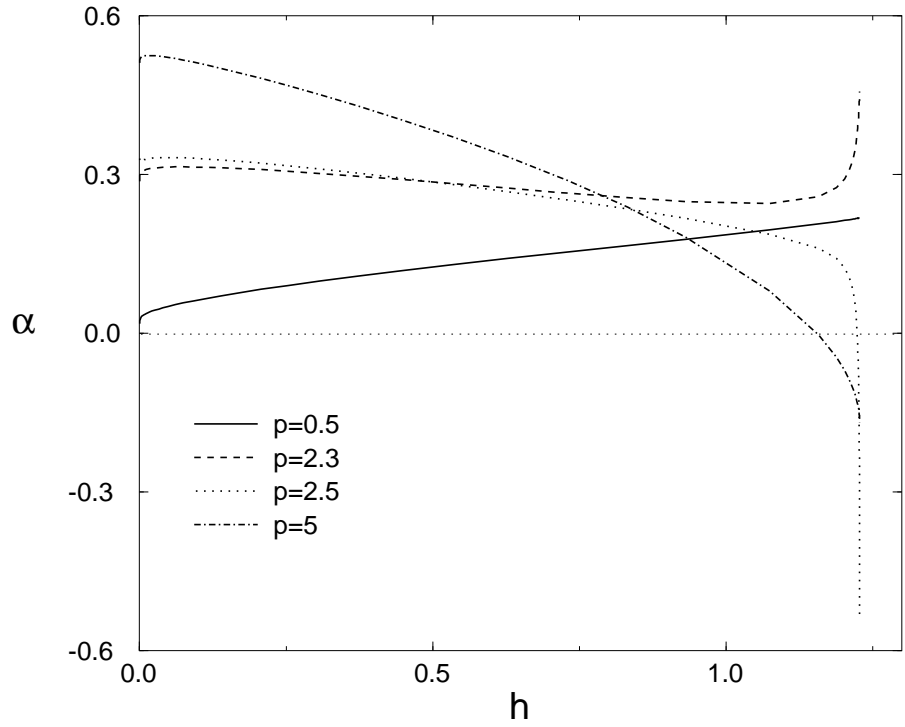}
\end{center}
\caption{
\label{fig:alphaH}
Magnetic field dependence of the edge exponent $\alpha_{abs}$ for
$u=1$ and fixed chemical potential $\mu=-0.2$ for several values of
the boundary potential $p$ above and below the threshold
$p_1=1+\sqrt{2}$ for creation of the holon/spinon bound state.}
\end{figure}

For vanishing magnetic field one has $\lambda^{(0)}=\infty$ allowing
to solve the spin part of the integral equations by Fourier
transformation.  As a consequence the dressed charge matrix ${\bf \sf
Z}$ (\ref{eq:z}) is function of a single variable $\xi=\xi(k^{(0)})$
\cite{woyn:89}, which is defined by the following integral equation:
\begin{equation}
  \xi(k)=1+\int_{-k^{(0)}}^{k^{(0)}} 
  {\rm d}k' \cos k'\ G(\sin k-\sin k') \xi(k')\ ,
\label{eq:xi}
\end{equation}
where $G(\lambda)={1 \over 4 \pi u}\mbox{Re}\left\{\Psi\left(1+i
{\lambda \over 4 u}\right)-\Psi\left({1 \over 2}+i {\lambda \over 4
u}\right)\right\}$ ($\Psi(x)$ is the digamma function).  Furthermore,
one finds $\theta^s_p={1\over2}\theta^c_p$ yielding
\begin{equation}
  \alpha_{abs}=1-2(x_c+x_s)={1 \over 2}-{1 \over \xi^2} 
  \left(1+\theta_{0}^c-\theta_{p}^c \right)^2 \ 
\end{equation}
for the exponent at the absolute threshold.
In Fig.~\ref{fig:h0a} we present regions where the exponent
$\alpha_{abs}$ is positive as a function of electronic density $n_e$
and boundary potential $p$ together with the density dependence of the
exponent for some  fixed values of $p$.
In a finite magnetic field $h$ the bulk state of the Hubbard model is
ferromagnetic below a critical particle density $n_c$.  This density
can be calculated from
\begin{equation}
  h={2u \over \pi} \int_{-\pi n_c}^{\pi n_c} {\rm d}k'\ \cos k'
  \frac{\cos k'-\cos(\pi n_c)}{u^2+\sin^2 k'}\ .
\label{eq:hc}
\end{equation}
For non vanishing magnetic field we will only consider electron densities 
above $n_c$.  For $h>0$ the exponent is given by:
\begin{equation}  
\alpha_{abs}=1-\frac{  
	         \left[\left(1+\theta^c_{0}-\theta^c_p\right)Z_{ss}
		- \left(\theta^s_{0}-\theta^s_p\right)Z_{cs}\right]^2
		+\left[\left(\theta^s_{0}-\theta^s_p\right)Z_{cc}
		- \left(1+\theta^c_{0}-\theta^c_p\right)Z_{sc}\right]^2}
	{\det^2({\bf \sf Z})} \ .
\end{equation}
In Fig.~\ref{fig:alphaH} the magnetic field dependence of this
exponent is shown for several values boundary potentials $p$.  Note
the characteristic change of this curve near $p=p_1=u+\sqrt{u^2+1}$
where a low lying excited bound state for a charge \emph{and} a spinon
(corresponding to a complex quasi momentum $k$ \emph{and} a complex
spin rapidity $\lambda$ in the set of roots of (\ref{eq:bae})
\cite{befr:97}) appears, see Fig.~\ref{fig:enpm}.
For $h$ approaching the saturation field (\ref{eq:hc}) the exponents
can be calculated explicitly \cite{bed:diss} giving
\[
  \alpha_{abs}(n_e \to n_c(h)) > 0 \quad \mbox{for} \quad 0<p<p_1 \qquad 
  \mbox{and} \quad 
  \alpha_{abs}(n_e \to n_c(h)) < 0 \quad \mbox{for} \quad p>p_1\ .
\]
The difference of the limiting values at $p=p_1$ is $\Delta
\alpha_{abs}=1$.  The crossover due to this behaviour is clearly seen
in Fig.~\ref{fig:alphaH}.
As before additional edge singularities arise as a consequence of
bound states in the boundary potential.  The corresponding exponents
can be computed from the Bethe Ansatz equations as above (for details
see Ref.~\cite{befr:97}).  Each of the boundary states seen in
Fig.~\ref{fig:enpm} gives rise to a singularity (\ref{eq:ea}) in the
$I$--$V$ curve.
\section{Summary and Conclusion}
We have studied the $I$--$V$ characteristics for tunneling from a
resonant localized level into a one dimensional interacting electron
gas described in terms of a Hubbard model (\ref{ham:hubb}).  Compared
to tunneling into a higher dimensional system one finds a rich
structure of thresholds due to the presence of various bound states in
the many particle spectrum, each of them leading to a possible
resonance for the tunneling charge.  Their appearance may be
understood as a consequence of the separation of charge and spin
degrees of freedom of the electrons allowing to generate a current
either by the electrons decaying into their holon and spinon
constituents in the wire \emph{or} by holons alone leaving the spinon
bound by the electrostatic potential of the tunnel contact.
Furthermore, the electronic correlations within the quantum wire
strongly affect the nature of the singularities, i.e.\ the exponents
$\alpha$ in (\ref{eq:ea}).  While we have considered correlation
functions strictly at zero temperature, the above singularities are
still observable at small finite $T$: for sufficiently long wires the
current at the threshold depends on temperature as $I(E_{th})\propto
T^{-\alpha}$.
Hence, the dependence of $\alpha$ on parameters such as filling factor
of the wire or the external magnetic field should be accessible
experimentally thereby allowing to determine the properties of the
potential due to a vacancy of the localized level.

\section*{Acknowledgments}
We are grateful to R.~Haug for discussions.  This work has been
supported by the Deutsche Forschungsgemeinschaft under Grant No.\
Fr~737/2--3.

\newpage

\setlength{\baselineskip}{16pt}

\newpage
\section*{Figures}


\end{document}